\documentclass[10pt, conference]{IEEEtran}
\usepackage[margin=1in,nohead,nofoot]{geometry}
\usepackage{subfigure}
\usepackage{epsfig,graphicx,psfrag}
\usepackage{amsfonts,amsmath,amssymb}
\usepackage{tikz}

\usetikzlibrary{arrows,positioning,fit}
\usetikzlibrary{calc}

\tikzset{arw/.style={->,>=stealth'}}

\newtheorem{theorem}{Theorem}[section]

\begin{document}

\title{Optimal Equivocation in Secrecy Systems a Special Case of Distortion-based Characterization}

\author{
\authorblockN{Paul Cuff -- Princeton University}
}

\maketitle

\begin{abstract}
Recent work characterizing the optimal performance of secrecy systems has made use of a distortion-like metric for partial secrecy as a replacement for the more traditional metric of equivocation.  In this work we use the log-loss function to show that the optimal performance limits characterized by equivocation are, in fact, special cases of distortion-based counterparts.  This observation illuminates why equivocation doesn't tell the whole story of secrecy.  It also justifies the causal-disclosure framework for secrecy (past source symbols and actions revealed to the eavesdropper).
\end{abstract}


Secrecy systems have a dual purpose---to reliably communicate to an intended receiver while hiding the information from an eavesdropper.  Reliable communication is typically measured operationally---i.e. the reconstruction of the information content must have a negligible probability of error, or when perfect transmission is not possible, a distortion constraint is met.  On the other hand, the conventional approach for information theoretic secrecy is to measure the secrecy aspect of the communication using equivocation (the posterior entropy of the information signal), rather than an operational quantity.  In this case, entropy and mutual information are a part of the problem statement itself, rather than being unexpectedly discovered in the answer.  Thus, this line of research forfeits one of the most elegant trademarks of information theory.

There are two regimes of interest in information theoretic secrecy, the perfect secrecy regime, and the partial secrecy regime.  The perfect secrecy case, where no information is leaked to the eavesdropper, is where the subtleties of ``strong secrecy'' versus ``weak secrecy'' (whether or not the leaked mutual information is negligible before or after being normalized by $n$) enters the conversation.  That is not the regime of interest in this paper.  This paper is focussed mainly on the partial secrecy case.

The use of equivocation ``as a measure of the degree to which the wire-tapper is confused'' can be found at least as early as Wyner's work on the wiretap channel \cite{wyner}.  A direct reference to this standard was made by Csisz\'{a}r and K\"{o}rner in their related work in \cite{csiszar-korner}, where they also use normalized equivocation to quantify partial secrecy.  This trend continues today, as is evident in these recent contributions:  \cite{villard-piantanida}, \cite{gunduz-erkip-poor}, \cite{lai-el-gamal}, \cite{tandon-ulukus-ramchandran}.  It is true that Shannon's much earlier 1949 paper on secrecy systems \cite{shannon} also refers to equivocation, but in his case it was simply used as an analysis tool, not a definition of secrecy---the extreme values of equivocation correspond to perfect secrecy and error-free decoding, both of which can be defined with simple, operational statements.

In \cite{cuff1} and \cite{cuff2}, we proposed a rate-distortion approach to measuring partial secrecy.  The communication system is designed to guarantee a certain level of distortion in the eavesdropper's reconstruction of the information source.  It was discovered that the richest theories emerge when the eavesdropper is empowered by having causal access to either the information signal or the intended receiver's reconstruction or both in addition to eavesdropping on the communication.  Without this causal disclosure of information, the communication system can get away with inexpensive tricks to keep the eavesdropper's distortion high even while the eavesdropper has very little uncertainty about the information signal \cite{schieler-cuff}.

The causal disclosure of information to the eavesdropper in \cite{cuff1} and \cite{cuff2} plays a fundamental role in the results, yet it is an uncommon assumption in communication systems.  An operational justification for causal disclosure is that the eavesdropper is actively trying to disturb the system---the distortion function measuring the eavesdropper's performance can be generalized to be the payoff of a zero-sum repeated game between the eavesdropper and the system.  In a repeated game context, disclosure of the past actions in the game can be a natural assumption.  However, the results of this work provide an entirely different justification for the rate-distortion-with-causal-disclosure framework for measuring partial secrecy.

It turns out that the equivocation approach to measuring partial secrecy and the rate-distortion approach are not in fact two different approaches.  The rate-distortion approach provides the general theory, yielding the optimal equivocation characterization as a special case.  In fact, equivocation is one of the most blunt special cases of the theory in the sense that it requires the least care in designing the optimal communication system.  The idea of this work is very simple.  We apply the theorems for rate-distortion theory with causal disclosure in \cite{cuff2} to the log-loss distortion function.  This technique immediately generalize to any secrecy setting.  That is, once distortion-based secrecy with causal disclosure is understood for a secrecy setting, then the characterization of optimal equivocation can be readily obtained.

\section{Rate-distortion theory with causal disclosure}

Consider a cipher system with an i.i.d. information source, $X_i \sim P_X$.  Secret key is available at a rate of $R_0$ bits per source symbol to the encoder $A$ and the decoder $B$.  The encoder transmits a message $M$ at rate $R$ bits per source symbol to the decoder which is also received by the eavesdropper.  The decoder then produces a sequence $Y_i$, the eavesdropper produces a sequence $Z_i$, and the performance of the system is measured by a payoff function $\pi(X_i,Y_i,Z_i)$

From \cite{cuff2} we obtain theorems that give so-call ``single-letter'' information theoretic characterizations of the following three quantities for any payoff function $\pi$ and source distribution $P_X$.
\begin{eqnarray}
  \sup_{A,B,n} \; \min_{ \{ Z_i = z_i(M,X^{i-1}) \} } \mathbf{E} \; \frac{1}{n} \sum_{i=1}^n \pi(X_i,Y_i,Z_i), \label{eq:  disclose x} \\
  \sup_{A,B,n} \; \min_{ \{ Z_i = z_i(M,Y^{i-1}) \} } \mathbf{E} \; \frac{1}{n} \sum_{i=1}^n \pi(X_i,Y_i,Z_i), \label{eq:  disclose y} \\
  \sup_{A,B,n} \; \min_{ \{ Z_i = z_i(M,X^{i-1},Y^{i-1}) \} } \mathbf{E} \; \frac{1}{n} \sum_{i=1}^n \pi(X_i,Y_i,Z_i).  \label{eq:  disclose x and y}
\end{eqnarray}
The difference between these three quantities is the information that the Eavesdropper has available when forming $Z_i$.

\section{Log-loss}

Consider three log-loss payoff functions
\begin{eqnarray*}
  \pi_1 (x,y,z) & = & \log \frac{1}{z(x)}, \\
  \pi_2 (x,y,z) & = & \log \frac{1}{z(y)}, \\
  \pi_3 (x,y,z) & = & \log \frac{1}{z(x,y)},
\end{eqnarray*}
where $x$ and $y$ have domains ${\cal X}$ and ${\cal Y}$, and $z$ is a probability mass function on the space ${\cal X}$ for $\pi_1$, the space ${\cal Y}$ for $\pi_2$, and the space ${\cal X} \times {\cal Y}$ for $\pi_3$.  In other words, $z(a) \geq 0$ for all $a$, and $\sum_a z(a) = 1$.

Log-loss functions have been studied as distortion functions for multiterminal source coding in \cite{courtade-wesel} and \cite{courtade-weissman}, in which the authors refer to the use of log-loss functions in the study of information bottleneck in \cite{harremoes-tishby} and as a distortion measure in image processing \cite{andre-antonini-barlaud-gray}.  Our reason for analyzing communication under this payoff (or ``distortion'') function is the following convenient property:
\begin{eqnarray*}
  \min_{Z = z(U)} \mathbf{E} \; \pi_1(X,Y,Z) & = & \min_{Z = z(U)} \mathbf{E} \; \log \frac{1}{Z(X)} \\
  & = & H(X|U),
\end{eqnarray*}
where $X$, $Y$, and $U$ are arbitrarily correlated.  Likewise,
\begin{eqnarray*}
  \min_{Z = z(U)} \mathbf{E} \; \pi_2(X,Y,Z) & = & H(Y|U), \\
  \min_{Z = z(U)} \mathbf{E} \; \pi_3(X,Y,Z) & = & H(X,Y|U).
\end{eqnarray*}

\section{Equivocation as a Special Case}
We now specialize the rate-distortion theory for secrecy systems with causal disclosure to the case of log-loss payoff functions.  For example, consider substituting $\pi_1$ as the payoff function in \eqref{eq:  disclose x}.
\begin{eqnarray*}
  & & \sup_{A,B,n} \; \min_{ \{ Z_i = z_i(M,X^{i-1}) \} } \mathbf{E} \; \frac{1}{n} \sum_{i=1}^n \pi_1(X_i,Y_i,Z_i) \\
  & = & \sup_{A,B,n} \; \min_{ \{ Z_i = z_i(M,X^{i-1}) \} } \mathbf{E} \; \frac{1}{n} \sum_{i=1}^n \log \frac{1}{Z_i(X_i)} \\
  & = & \sup_{A,B,n} \; \frac{1}{n} \sum_{i=1}^n H(X_i|M,X^{i-1}) \\
  & = & \sup_{A,B,n} \; \frac{1}{n} H(X^n|M).
\end{eqnarray*}
This quantity defines the optimal normalized equivocation (with respect to the information source $X$) that a cipher can achieve.  We can then apply the information theoretic theorems of \cite{cuff2} to obtain a characterization of this quantity---the optimal equivocation.

Before we proceed we must point out one subtle detail.  The log-loss payoff function puts no constraint that the intended receiver must produce an output $Y$ that is correlated with the information source $X$.  In fact, it rewards the system for choosing $Y$ independent of $X$ and sending no description of $X$.  In order to impose a genuine communication requirement, we add a second distortion constraint, aside from the log-loss payoff.  The constraint is $\frac{1}{n} \sum_{i=1}^n d(X_i,Y_i) < D$, and the theorems of \cite{cuff2} are readily generalized to include more than one payoff (or distortion) function.

This leads us to our resulting statements about equivocation in a cipher system.  Upon simplifying Theorems 7.1, 7.2, and 3.1 of \cite{cuff2} using the payoff functions $\pi_1$, $\pi_2$, and $\pi_3$ we obtain the following theorem (proof of simplification omitted):

\begin{theorem}
  For a cipher system with communication rate $R$, secret key rate $R_0$, source distribution $P_X$, and satisfying an average distortion constraint of $D$ at the intended receiver (as measured by $d(x,y)$), the following three statements hold:
  
  \noindent
  Equivocation of the information source:
  \begin{eqnarray*}
    \sup_{A,B,n} \; \frac{1}{n} H(X^n|M) & = & H(X) - [R(D) - R_0]_{+}.
  \end{eqnarray*}
  
  \noindent
  Equivocation of the reconstruction:
  \begin{eqnarray*}
    & & \sup_{A,B,n} \; \frac{1}{n} H(Y^n|M) \\
    & = & \max_{(X,Y,U) \in {\cal P}} \left( H(Y) - [I(Y;U) - R_0]_{+} \right).
  \end{eqnarray*}
  
  \noindent
  Equivocation of the source and reconstruction:
  \begin{eqnarray*}
    & & \sup_{A,B,n} \; \frac{1}{n} H(X^n,Y^n|M) \\
    & = & \max_{(X,Y,U) \in {\cal P}} \left( H(X,Y) - [I(X,Y;U) - R_0]_{+} \right).
  \end{eqnarray*}
  Here $[\cdot]_{+} = \max \{ 0, \cdot \}$, $R(D)$ is the rate-distortion function for distortion $d$ and source distribution $P_X$, and ${\cal P}$ is the set of random variable triples that satisfy
  \begin{eqnarray*}
    X \;\; - & U & - \;\; Y \; \mbox{  form a Markov chain,} \\
    \mathbf{E} \; d(X,Y) & \leq & D, \\
    I(X;Y) & \leq & R.
  \end{eqnarray*}
\end{theorem}

\section{Summary}
Information theory allows us to understand the optimal use of communication resources not only in the ideal settings of lossless compression, noise free communication channels, or perfect secrecy, but also in the imperfect settings.  Rate-distortion theory helps us understand optimal partial communication.  Similarly, we endeavor to understand partial secrecy, for situations that do not permit perfect secrecy.  We see from this work that a distortion-based measure of secrecy which includes causal disclosure of information to the eavesdropper (in addition to the intercepted communication signal) provides a rich theory for understanding secrecy.  This theory subsumes the conventional equivocation based approach to secrecy in that the latter is a special case of the rate-distortion theory via the log-loss function.

This analysis also highlights the blunt nature of the equivocation approach to secrecy.  In general, distortion-based performance in a secrecy system is optimized by a carefully crafted communication scheme.  However, in the special case of log-loss (i.e. the equivocation characterization), many different encoding variations are optimal, including in particular random binning or time-sharing.  An analogy to this occurs in lossy compression.  If the goal in lossy compression were simply to minimize the equivocation at the intended receiver, then any message transmission would be equally good, as long as it is a fully entropic function of the information source.  But rate-distortion theory reveals much more than this about the nature of partial information.

\section{Acknowledgment}
The author would like to thank Tsachy Weissman for suggesting the study of the log-loss function as a special case.  This work is supported by the National Science Foundation through grant CCF-1116013 and by the Air Force Office of Scientific Research through grant FA9550-12-1-0196.

\end{document}